# On the origin of diverse interlayer charge redistribution in transition-metal dichalcogenides

Yu-Meng Gao, Nie-Wei Wang, Shi-Xuan Yuan, Wen-Xin Xia, Jiang-Long Wang*, and Xing-Qiang Shi*

Key Laboratory of Optic-Electronic Information and Materials of Hebei Province, Hebei Research Center of the Basic Discipline for Computational Physics, College of Physics Science and Technology, Hebei University, Baoding 071002, P. R. China

*E-mails: jlwang@hbu.edu.cn, shixq20hbu@hbu.edu.cn

**Abstract**: The interlayer quasi-chemical-bonding (QCB) interactions of two-dimensional (2D) layered materials promote the research field of interlayer-engineering and cause interlayer charge density redistributions (ICDRs). The ICDRs have been reported experimentally and theoretically, which show different redistributions, e.g., accumulation, depletion, or a more complicated behavior. The underlying mechanism for the different ICDRs remain to be elucidated. In the current work, via a systematic theoretical study of the ICDRs of transition metal dichalcogenides with different number of $d$-electrons filling ($d^0$ TiS$_2$, $d^1$ NbS$_2$, and $d^2$ MoS$_2$) in T and H phases, we reveal three mechanisms based on the coexistence of different types of interlayer QCB interactions. Mechanism (1) is from a competition between two types of interlayer interactions: namely, the interlayer interaction between *fully occupied* energy levels (in short: *o-o* interaction) depletes electrons in the overlap region while that between occupied and empty levels (*o-e* interaction) promotes electron accumulation; and the competition between them leads to that the $d^0$ TiS$_2$ tends to electron accumulation in T phase than in H phase. Mechanism (2), the interlayer interaction between half-filled levels (*h-h* interaction) promotes the electron accumulation of $d^1$ NbS$_2$. Mechanism (3), the interlayer interaction of multiple filled-levels of $d^2$ MoS$_2$ (namely, the multi-level *o-o* interaction) leads to a more complicated ICDR. The current study provides a unified understanding to the different ICDRs of van der Waals materials and paves the way for further exploration of their electronic properties and applications.

# I. INTRODUCTION

Layered van der Waals (vdW) materials have sparked substantial interest for both fundamental research and potential applications [1-6]. Transition metal dichalcogenides (TMDs) have been widely studied due to their excellent electronic properties and diverse applications [7-12]. TMDs generally have different structural phases, such as the T and H phases [13, 14]. At the interlayer region of TMDs, in addition to the dispersion attraction between layers, there also exist interlayer quasi-chemical-bonding (QCB) interactions, which can modulate the electronic and magnetic structure, lattice vibration, microscopic friction, and Schottky barrier height [15-20], to name a few. In addition to the energy level splitting, the interlayer interaction also leads to the interlayer charge density redistribution (ICDR) [16, 21-23]. For the study of ICDR of TMDs, Zhao *et al.* conducted a combined experimental and theoretical study on $PtS_2$ and $MoS_2$ and found that the interlayer S⋯S QCB of the T phase $PtS_2$ exhibiting a covalent character with electron accumulation in the interlayer region, while for the H phase $MoS_2$ the charge density redistribution in the interlayer region is less apparent [21]. Also for the H phase $MoS_2$, density-functional theory (DFT) calculations with the B3LYP hybrid functional reported an electron depletion between $MoS_2$ layers [24], which meets the physical picture of Pauli repulsion between fully occupied orbitals (or the two-level four-electron repulsion) [25]. Iversen *et al.* used synchrotron X-ray diffraction to study the T phase $TiS_2$ and reported electron accumulation between interlayer S⋯S [22]. Also, for T phase $TiS_2$, DFT calculations with some density functionals can match the experimental result of interlayer electron accumulation [26].

Interlayer QCB interactions can be classified into three main categories based on the energy changes induced by QCB, and some of them are summarized in literature [15]. The three main categories of interlayer QCB interaction are: (1) with energy cost, such as interactions between fully occupied energy levels; (2) does not change energy, such as interlayer interactions between unoccupied levels; (3) with energy gain, such as interlayer interaction between occupied and unoccupied levels, or, between half-filled levels. At present, the different interlayer QCB interactions of two-dimensional (2D) materials has been studied for energy-related changes/band edge energy positions (but not for charge density changes), either from perspectives of the above-mentioned different categories of interlayer interactions, or from a perspective of interlayer multi-level interaction, including interlayer interaction between multiple filled-levels or a combination of the above-mentioned different categories of interlayer interactions [24, 27-29]. For the charge density changes due to interlayer QCB, there is still a lack of understanding for the different ICDRs (tends to electron accumulation, depletion or more complicated redistributions at the middle region of van der Waals gap) for TMDs with transition metals of different groups or in T and H phases from the perspective of different categories of interlayer interactions and interlayer multi-level interactions.

In the current work, based on DFT calculations and a two-center two-level model as well as its extension to multi-levels, we develop a unified understanding to the different ICDRs of TMDs with different number of *d*-electrons filling ($d^0$ to $d^2$) and with T or H phase. We find that for the ICDR of $d^0$ TiS$_2$, the T phase tends to accumulate electrons relative to the H phase. Because there is a competition between two types of interlayer interactions ---- interlayer occupied-occupied and occupied-empty interactions. The former depletes electrons in the interlayer region owing to its net antibonding character, whereas the latter accumulates electrons for its bonding character. The interlayer occupied-empty interaction is stronger in the T phase than that in the H phase: this leads to a tendency that the T phase tends to accumulate electrons than the H phase. For the $d^1$ NbS$_2$, it is more inclined to accumulate electrons in the interlayer region than the $d^0$ TiS$_2$, due to the interlayer interaction between half-filled levels leads to electron accumulation. For the $d^2$ MoS$_2$, under the interlayer interaction of multiple filled-levels, a more complex ICDR behavior appears. In summary, for TMDs with transition metals of different groups and in different structural phases, the transition metal atom is in different coordination fields and with different number of *d*-electrons filling, which is related to the various ICDR behaviors. Our findings provide a fundamental understanding to the various ICDR behaviors of vdW layered materials and pave the way for further exploration of their electronic properties for interlayer-engineering.

## II. CALCULATION METHODS

Density-functional theory [30] calculations were performed using the Vienna *ab initio* Simulation Package (VASP) [31, 32]. The projector augmented-wave (PAW) potentials were adopted to describe the core electrons [33, 34]. The valence electrons were described by plane-wave basis with an energy cut-off of 500 eV. The interlayer van der Waals (vdW) interactions were included by the DFT-D3 method of Grimme *et al.* [35]. The Brillouin zone was sampled using the Gamma-centered mesh with a k-point density of 2π×0.03 Å$^{-1}$. For few-layer, a vacuum of at least 15 Å along the z-axis was used to avoid interaction between periodic images of the slab model. The atomic positions were relaxed until the force on each atom was less than 0.02 eV/Å and the convergence criteria for energy were set to 10$^{-5}$ eV. The electronic structure calculations were carried out with the B3LYP hybrid density functional, where the AEXX parameter was set to 0.18 to achieve consistency between the calculated charge density redistribution and that reported in experiment [22, 36-39], and to correctly capture the "two-level four-electron repulsion interaction" from calculated charge density redistribution [24]. In addition, some other functionals can also reproduce phenomena consistent with the experimental observations [22], as illustrated in Fig. S1 of the Supplemental Material [40]. The VASPKIT program was used for the post-processing of the electronic structure data [42].

## III. RESULTS AND DISCUSSION

In order to analyze the ICDRs resulting from interlayer orbital coupling of TMDs, the two-center two-level model between two atoms [25] can be borrowed here for use, and we also extend it to the situation of multiple filled-levels. For the two-level interactions in Figs. 1(a-c), the overlap population ($P_{12}$) between two atomic orbitals ($\chi_1$, $\chi_2$) is used to characterize the electron density redistribution at the middle region [25]. For the two atomic orbitals discussed here, when applied to the interlayer interaction of vdW layered materials, they correspond to the orbitals from two adjacent layers separated by the vdW gap.

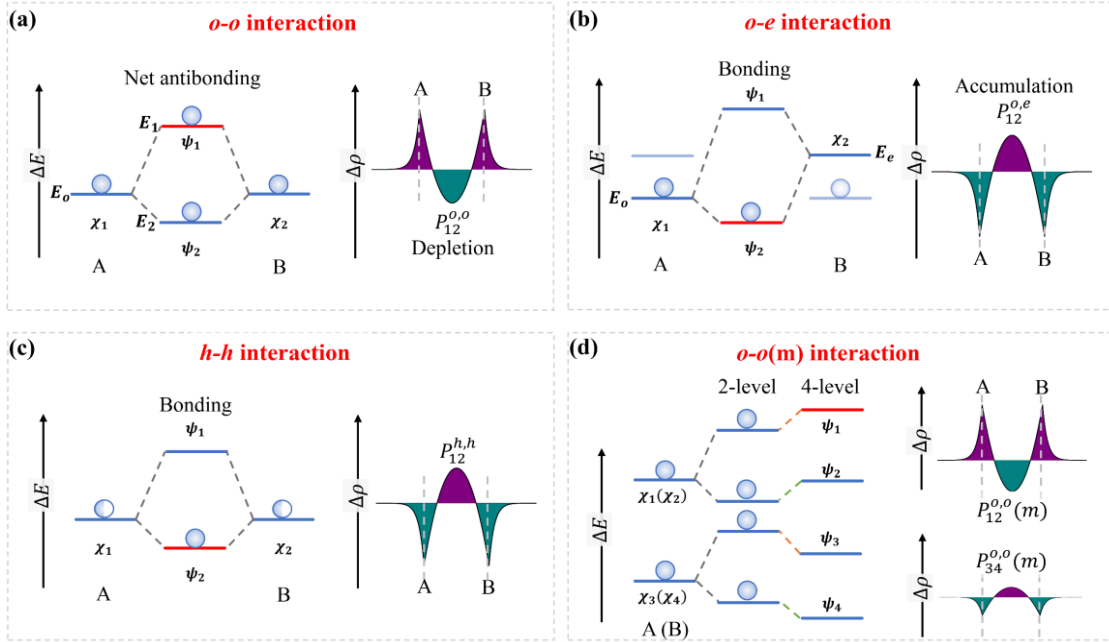

**FIG. 1. Schematic diagram of several types of interlayer interactions and the resulting different ICDRs as described by the various overlap populations ($P_{12}$).** (a) Interlayer interaction between occupied levels (*o-o* interaction), (b) between occupied and empty levels (*o-e* interaction), (c) between half-filled levels (*h-h* interaction), and (d) between multiple occupied levels [*o-o*(*m*) interaction]. In (a-d), the left panels sketch the energy level splitting with the overlap integral included (see text for details), in which the bilayer energy level in red represents the dominant level for the overall bonding or antibonding feature; the right panels sketch the corresponding ICDRs, with green (purple) indicating electron density depletion (accumulation). In the left panel of (d), the labeled "2-level" ("4-level") denotes interlayer two-level (four-level) interaction (see text). The various overlap populations [e.g., $P_{12}^{o,o}$, $P_{12}^{o,o}(m)$] are illustrate in the text. A filled (half-filled) circle on an energy level means fully (half) occupied level.

*Interlayer two-level interactions.* The energy splitting in the left panel of Fig. 1(a) shows that, the magnitude of the energy up-shift of the antibonding-level (relate to $\psi_1$) is larger than the energy down-shift of the bonding-level (relate to $\psi_2$), which is due to the overlap integral ($S$) between layers.

Without $S$ included, the energy up-shift (down-shift) of the antibonding-level (bonding-level) is $t$, where $t$ is the interlayer hopping intergral; with $S$ considered, the energy up-shift (down-shift) of the antibonding-level (bonding-level) is $\frac{t}{1-S}$ ($\frac{t}{1+S}$) [25]. Namely, the strength of anti-bonding is larger than bonding, and hence a net antibonding is obtained in Fig. 1(a) for the interlayer interaction of two occupied levels (the *o-o* interaction). The net antibonding character results in electron depletion at the middle region, as denoted by $P_{12}^{o,o}$ (the overlap population of interlayer *o-o* interaction) in the right panel of Fig. 1(a). More on overlap populations can be found in Note S1 of the Supplemental Material [40]. In Fig. 1(b), for the interlayer interaction between occupied and empty levels (the *o-e* interaction), only the bonding level is occupied, and hence the corresponding overlap population, $P_{12}^{o,e}$, is electron accumulation. For the interlayer *o-e* interaction, each layer contributes an occupied level and an empty level, and, for clarity, only the interlayer interaction between the occupied level on the left and the empty level on the right is shown in Fig. 1(b). The ICDR of some vdW materials (like the $d^0$ TiS$_2$) can be seen as the result of a competition between $P_{12}^{o,o}$ and $P_{12}^{o,e}$, as illustrated in subsection A below. In Fig. 1(c), the interlayer interaction between half-filled levels (the *h-h* interaction) also leads to electron accumulation ($P_{12}^{h,h}$), like $P_{12}^{o,e}$ in Fig. 1(b) with electron accumulation. However, $P_{12}^{h,h}$ may be larger than $P_{12}^{o,e}$, since the former (latter) is from the interaction of degenerate (nondegenerate) energy levels, [an example of this will be introduced in subsection B for the $d^1$ NbS$_2$].

*Interlayer four-level interaction.* In Fig. 1(d), we further extend the two-level interlayer interaction in Fig. 1(a) to multiple filled-levels in each layer [the *o-o*(m) interaction], here two-filled levels from each layer with the same orbital character. The two levels in each layer may come from, for example, the *intralayer p-d* hybridization, so that the $p_z$-orbital contribute to more than one occupied energy levels [24]. Although the $\chi_1$ and $\chi_3$ within a single layer (layer A) are orthogonal and hence do not interact any more, they can interact with the $\chi_2$ and $\chi_4$ of layer B. In Fig. 1(d) with more than one filled-levels each layer, the interlayer two-level interaction means $\chi_1$ interacts with $\chi_2$ and $\chi_3$ interacts with $\chi_4$; and, the interlayer four-level interaction means $\chi_1$ interacts with $\chi_2$ and $\chi_4$, $\chi_3$ interacts with $\chi_4$ and $\chi_2$. The four-level interaction pushes energy-levels further move up (move down) for the upper (lower) levels relative to the two-level interaction, and the move up (move down) of energy levels corresponds to a change towards the trend of antibonding (bonding) [43]. As a result, under the four-level interaction, the energy levels corresponding to $\psi_1$, $\psi_2$ ($\psi_3$, $\psi_4$) are up-shifted (down-shifted) compared to the two-level interaction, which leads to: 1) the enhanced electron-depletion of $P_{12}^{o,o}(m)$ relative to $P_{12}^{o,o}$ ($m$ denotes interlayer multi-level interaction), and 2) the inversed $P_{34}^{o,o}(m)$ from electron depletion to electron-accumulation (see Note S2 and Fig. S4 in the Supplemental Material [40]). Then, $P_{34}^{o,o}(m)$ from the lower energy-levels becomes electron accumulation due to the four-level interaction, while its absolute value is smaller than $P_{12}^{o,o}(m)$. The net ICDR for the four-level interaction is: $P_{12}^{o,o}(m) - P_{34}^{o,o}(m)$, which leads a more complex ICDR behavior [which will be introduced in subsection C].

### A. The competition between *o-o* and *o-e* interactions

The T and H structural phases are commonly found for TMDs. In the T (H) phase, the metal atoms adopt an octahedral (trigonal prismatic) coordination as illustrated in Fig. S2, and the calculated energy difference between the two phases for the TMDs investigated in this work is summarized in Table SI of the Supplemental Material [40]. The optimized lattice constant of the T phase bulk $TiS_2$ is $a$ = 3.39 Å with an interlayer separation of 5.70 Å in the $c$ direction, which is consistent with the literature [22], and that of the H phase is $a$ = 3.34 Å with an interlayer separation of 5.95 Å. Figure 2 shows the electron density redistribution along the $c$ direction (or $z$-direction), $\Delta\rho(z)$, for T and H phases bulk $TiS_2$ compared to the free-standing monolayers.

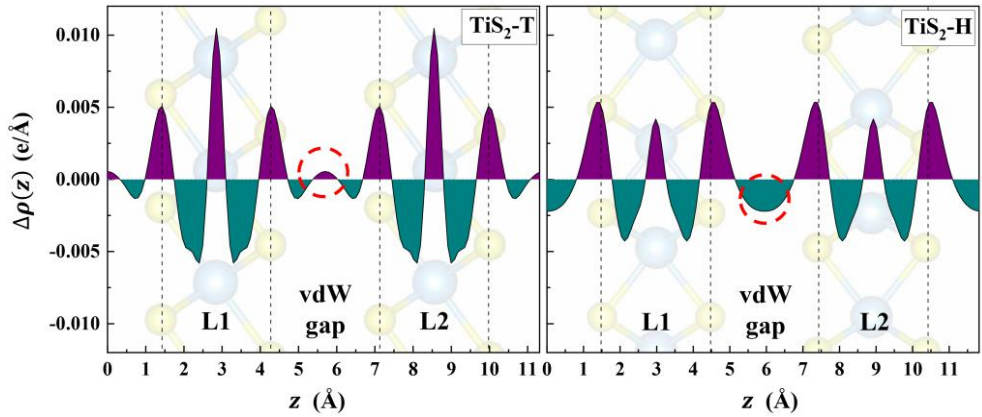

**FIG. 2.** ICDR from the plane-averaged electron density redistribution along the $z$-direction, $\Delta\rho(z) = \rho(\text{bulk}) - \rho(\text{monolayers})$, for T phase (left) and H phase $TiS_2$ (right). Purple (green) represents electron accumulation (depletion). The circle in each panel highlights the electron density redistribution at the middle of vdW gap.

At the middle region of the vdW gap between $TiS_2$ layers, the electron density redistribution of T phase $TiS_2$ exhibits a trend of electron accumulation, which is consistent with the conclusion from X-ray electron density experiment [22]. In contrast, the H phase shows electron depletion at the middle region. Integrated electron accumulation/depletion at the middle of the van der Waals gap for T and H phases $TiS_2$ are summarized as shown in Table SII of the Supplemental Material [40]. To understand the different trends of electron accumulation/depletion in the vdW gap for T and H phases, a comparison of their interlayer QCB interactions is helpful. We start from the real-space wavefunctions of monolayer to study the interlayer interaction of a bilayer. To analyze the interlayer interactions of $TiS_2$, the band structure is projected to the out-of-plane $p_z$ orbital of S atoms and $d_{z^2}$ orbital of Ti atoms for both phases (Fig. 3), because of the interlayer interaction is along the out-of-plane z-direction. We find that the energy level splitting at the Γ point is significant from monolayer to bilayer (Fig. S3 of the Supplemental Material [40]); and correspondingly, the $p_z+d_{z^2}$ orbital weight is significant. Therefore, we focus on the energy levels at the Γ point to study interlayer QCB interaction.

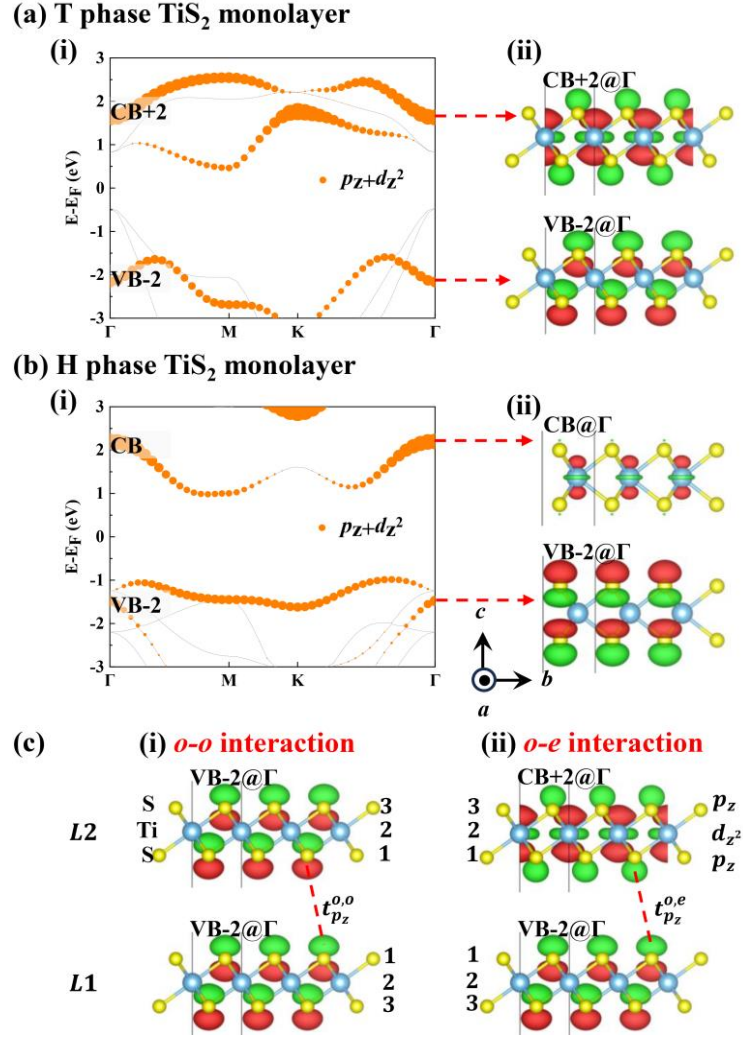

FIG. 3. **Projected band structures and real-space wavefunctions of the T and H phases TiS$_2$ monolayers, and the interlayer o-o and o-e interactions.** (a)(i) and (b)(i) show the band structures projected to the out-of-plane $p_z$ and $d_{z^2}$ orbitals; the dot size is proportional to the magnitude of projection. (a)(ii) and (b)(ii) display the wavefunctions at the Γ point for the occupied and empty levels with significant $p_z$ and/or $d_{z^2}$ contributions. The isosurface value is $5.4 \times 10^{-7}$ Bohr$^{-3/2}$ for T phase and $9.0 \times 10^{-7}$ Bohr$^{-3/2}$ for H phase. (c) Sketches for the interlayer o-o interaction ($t_{p_z}^{o,o}$) and o-e interaction ($t_{p_z}^{o,e}$), taking T phase as an example.

For the T phase in Fig. 3(a)(i), the projections on $p_z$ and $d_{z^2}$ orbitals are pronounced for bands VB-2 and CB+2 at the Γ point; and the corresponding real-space wavefunctions are shown in Fig. 3(a)(ii). The wavefunctions shows that the VB-2@Γ is contributed by the $p_z$ orbital of upper and lower S atoms; while for CB+2@Γ, it is contributed by both the $d_{z^2}$ orbital of Ti and the $p_z$ orbitals of S atoms. The $p_z$-orbital weights are close to each other in CB+2@Γ and VB-2@Γ. In CB+2@Γ, the inner-half of the $p_z$-orbital disappears due to the $p_z$-$d_{z^2}$ anti-bonding interaction. For the H phase, Fig. 3(b)(i) shows that the $p_z+d_{z^2}$ orbital weights at VB-2@Γ and CB@Γ are pronounced in the band structure. The corresponding wavefunctions are shown in Fig. 3(b)(ii), which shows that the VB-2@Γ is contributed by the $p_z$ orbitals of the sulfur atoms, similar to the T phase. However, for the unoccupied level CB@Γ, unlike the wavefunction of CB+2@Γ of T phase, the empty level wavefunction of H phase has a small contribution of $p_z$ orbital from S atoms compared to the $p_z$ orbitals of VB-2@Γ. For interlayer orbital interaction, since the interlayer Ti···S distance is about one Ångstrom larger than the interlayer S···S distance for both T and H phases (see Table SIII in the Supplemental Material [40]), the interlayer S···S $p_z - p_z$ orbital interaction dominates the interlayer QCB interaction. For TiS$_2$, the $p_z$ orbital appears in both the VB and CB, and both contribute to interlayer interactions. Namely, both the interlayer *o-o* and *o-e* interactions contribute to the ICDR. The *o-o* and *o-e* interactions are sketched in Fig. 3(c). The interlayer empty-empty (*e-e*) interaction is not considered since the unoccupied levels do not contribute to electron density. As a result, there is a competition between the interlayer *o-o* and *o-e* interactions for the ICDR, and recall the overlap populations in Figs. 1(a) and 1(b) for interlayer *o-o* and *o-e* interactions. For the T phase, the weight of the $p_z$ orbital of CB+2@Γ is almost equivalent to that of the $p_z$ orbital of VB-2@Γ, resulting in stronger interlayer *o-e* interaction. Whereas, for the H phase, the weight of the $p_z$ orbital of CB@Γ is smaller than that of the $p_z$ orbital of VB-2@Γ, resulting in weaker interlayer *o-e* interaction. This explains why the ICDR calculated by DFT in Fig. 2(a) shows a trend of electron accumulation at the middle of the vdW gap, while the H phase in Fig. 2(b) shows a significant electron-depletion in the vdW gap.

Through the above overlap populations, $P_{12}^{o,o}$ and $P_{12}^{o,e}$ in Figs. 1(a, b) and the corresponding Eqs. (S5) and (S6) in the Supplemental Material [40], the interlayer *o-o* interaction depletes electrons at the interlayer region ($P_{12}^{o,o}$), whereas interlayer *o-e* interaction accumulates electrons at the interlayer region ($P_{12}^{o,e}$). Therefore, for the ICDR of TiS$_2$, the T phase tends to accumulate electrons relative to the H phase, which is the result of the competition between interlayer *o-o* and *o-e* interactions. The above is corresponding to Figs. 1(a, b) for the $d^0$ TiS$_2$. Furthermore, Fig. 1(c) [Fig. 1(d)] applies to other systems such as $d^1$ NbS$_2$ [$d^2$ MoS$_2$], which will be discussed in the following subsections.

### B. Half-filled levels

In the above, we explained the difference ICDR behaviors between T and H phases in $d^0$ TiS$_2$ originate from the different combinations of *o-o* and *o-e* interactions: the former tends to accumulates electrons at the middle of the vdW gap, while the latter depletes electrons significantly at the vdW gap.

For the $d^1$ NbS$_2$, there is also a trend that the T phase tends to accumulate electrons compared to the H phase, as shown in Fig. 4.

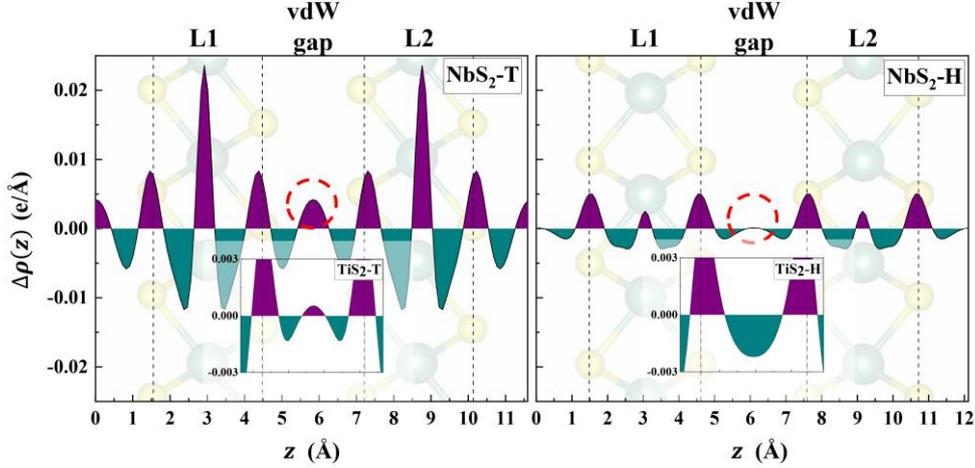

**FIG. 4.** ICDR of $d^1$ NbS$_2$. That for TiS$_2$ are also given in the insets for comparison. Purple (green) represents accumulation (depletion) of electrons.

From Fig. 4 and Table SII of the Supplemental Material [40], we can see that NbS$_2$ is more inclined to accumulates electrons in the vdW gap than TiS$_2$, for both T and H phases. This is because, for the metallic $d^1$ NbS$_2$, in addition to the interlayer interactions in $d^0$ TiS$_2$, there exists an interlayer interaction between half-filled levels (*h-h* interaction), which also leads to interlayer electron accumulation ($P_{12}^{h,h}$) as shown in Fig. 1(c). The electron accumulation intensity of the interaction between half-occupied levels (degenerate energy levels) can be stronger than that of the interaction between occupied and empty levels (nondegenerate energy levels). Additionally, our recent work indicated that the *h-h* interaction occurs in the Brillouin zone (BZ) at and around the k-point associated with the half-filled level [44]. Namely, electron transfer takes place within an area in the BZ but not an isolated k-point, and gives rise to electron accumulation in the vdW gap since, in this k-area, the bonding levels are occupied and antibonding levels are empty [44].

### C. Multiple filled-levels

For the $d^2$ MoS$_2$, for the interlayer electron density of MoS$_2$ in the T phase, Fig. 5(a) shows that it does not accumulate electrons as the above-discussed $d^0$ and $d^1$ TMDs in T phase, but deplete electrons and with a complex shape in the middle region. The depletion of electrons in the interlayer region originates from an additional *o-o* interaction in MoS$_2$ relative to TiS$_2$, due to the multiple-filled levels in MoS$_2$; and, the complex shape of electron density redistribution in the interlayer region of Fig. 5(a) requires further analysis.

In the two-level model of Fig. 1(a), we only consider a single occupied energy level. In $d^2$ MoS$_2$,

there are interlayer interaction between multiple filled energy levels, such as the interlayer six-level interaction of H phase MoS$_2$ [24].

Here, we extend our study from interlayer two-level interactions to interlayer four-level interaction, and the corresponding interaction is schematically shown in Fig. 1(d) and Fig. S4 in the Supplemental Material [40]. In Fig. 5, we consider two occupied energy levels each layer for the interlayer four-level interaction, namely, the mVB-4, and mVB-7 in Fig. 5(b) that mainly contributed by $p_z$ orbitals at the Γ point for the T phase monolayer MoS$_2$. The mVB-4 (mVB-7) levels from the two monolayers (L1 and L2) undergo interlayer interaction and split into bVB-6/bVB-7 (bVB-12/bVB-13) levels in the bilayer [Figs. 5(b-c)]. A cross-comparison of Figs. 5(b-c) and Fig. 1(d) reveals the following correspondences: mVB-4 in Fig. 5(b) corresponds to $\chi_1$ ($\chi_2$) in Fig. 1(d), bVB-6/bVB-7 in Fig. 5(c) corresponds to $\psi_1, \psi_2$; and mVB-7 corresponds to $\chi_3$ ($\chi_4$), bVB-12/bVB-13 corresponds to $\psi_3, \psi_4$. For the ICDR caused by the interlayer interactions of the mVB-4 and mVB-7 levels, we found that, in addition to the interlayer electron depletion corresponding to the mVB-4 energy level [Fig. 5(d), corresponding to the $P_{12}^{o,o}(m)$ in Fig. 1(d) or Fig. S4(a)], the mVB-7 energy level corresponds to the accumulation of interlayer electrons [Fig. 5(e), corresponding to the $P_{34}^{o,o}(m)$ in Fig. 1(d) or Fig. S4(b)]. The $\Delta\rho(z)$ in Fig. 5(d) is obtained as:

$$\Delta\rho(z) = \rho(\text{bVB-6}@\Gamma) + \rho(\text{bVB-7}@\Gamma) - \rho(\text{mVB-4}@\Gamma, \text{L1}) - \rho(\text{mVB-4}@\Gamma, \text{L2}). \quad (1)$$

The $\Delta\rho(z)$ in Fig. 5(e) is obtained in a similar way.

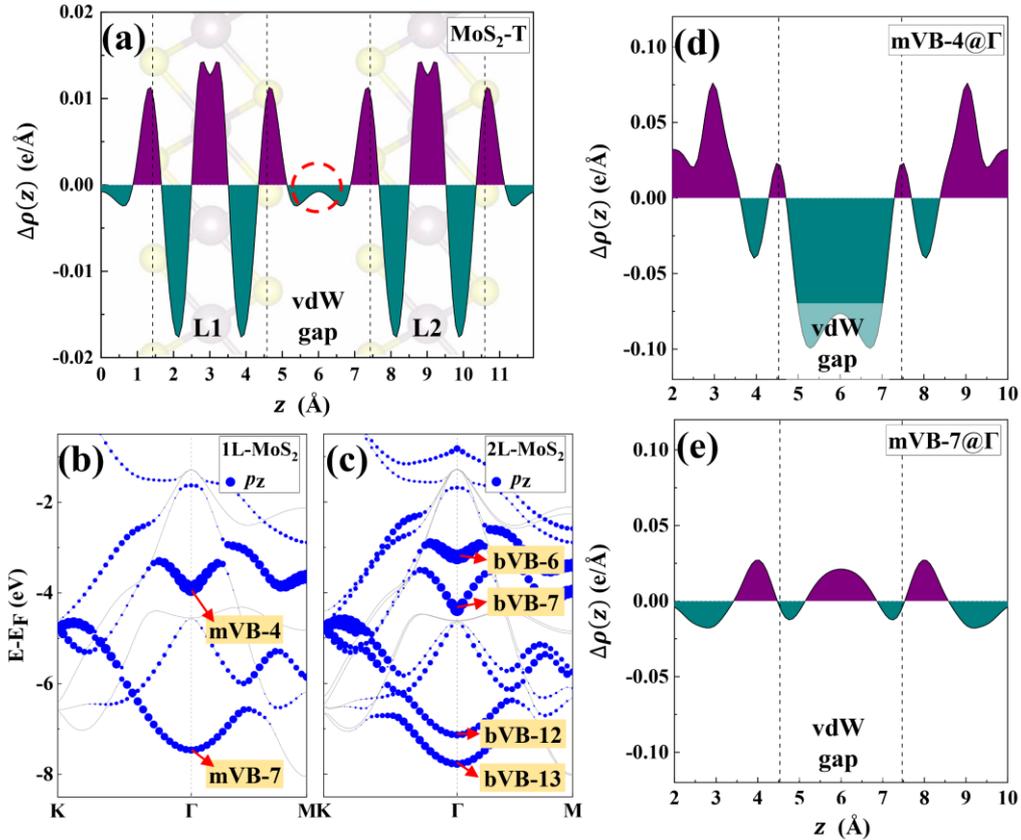

**FIG. 5. ICDR of $d^2$ MoS$_2$ in T phase.** (a) Electron density redistribution along the z-direction. Purple (green) represents accumulation (depletion) of electrons. Band structure projected to the out-of-plane $p_z$ orbital of monolayer (b) and bilayer (c); the dot size is proportional to the magnitude of projection. (d-e) Electron density redistribution of mVB-4@Γ and mVB-7@Γ.

Due to the interlayer interactions involving multiple occupied levels (here two levels each layer), the *o-o* interactions of multi-levels [*o-o(m)* interaction], both mVB-4 and mVB-7 contribute to electron redistribution, as given in Fig. 5(d) and Fig. 5(e), respectively. The competition between $P_{12}^{o,o}(m)$ and $P_{34}^{o,o}(m)$, here between the mVB-4 and mVB-7 derived bVB-6/bVB-7 and bVB-12/bVB-13, results in the complex shape in the overall ICDR of Fig. 5(a). For Fig. 5(a), it can be seem come from the energy level contributions of the entire Brillouin zone [refer to Figs. 5(b, c) for the K-Γ-M path], including the Γ point and other k-points. Our analysis focuses on the Γ point [Figs. 5(d, e)], where the energy level splitting is the most significant from monolayer to bilayer [Figs. 5(b, c)]. This provides an intuitive understanding of the phenomenon. In addition, the different energy levels in multi-level interaction (such as the mVB-4 and mVB-7 levels) may have different ratios of $p_z$, $d_{z^2}$ and *s*-orbitals, and this may lead even more complexity in ICDR.

Through systematic theoretical investigation of ICDRs in TMDs with different number of *d*-electrons filling and with T or H phases, we reveal the origins of different ICDRs based on the combination/competition of different types of interlayer-QCB interactions. Next, we examine the effect of structural phases and energy-level filling, providing insights into these different ICDRs in TMDs from another perspective.

### D. Effect of structural phases and energy-level filling

In addition to understanding the ICDR behaviors of different TMDs ($d^0$ TiS$_2$, $d^1$ NbS$_2$, and $d^2$ MoS$_2$) from the perspective of different interlayer interactions, it can give additional insights from the perspectives of their different crystal field splitting in T/H phases and *p-d* hybridized energy-levels with different number of *d*-electrons filling within a monolayer. Namely, beginning from an *intralayer* perspective.

The difference between the T and H structure phases is that the *d* orbital is under different coordination fields, which causes different energy level splitting. In the Fig. 6(a), the *d*-levels in the T phase split into two groups ($e_g$ and $t_{2g}$) under the octahedral field. For the H phase, under the triangular prism coordination, the *d* orbitals split into three groups, two doublets and one singlet. As a result of the different *d*-orbital splitting in T and H phases, the energy difference between *p* and lowest *d* level(s) is smaller in the H phase than that in the T phase [Fig. 6(a)]. Then, the intralayer *p-d* hybridization is stronger in the H phase and the energy splitting from *p-d* hybridization is larger in the H phase, which means larger energy gaps of $d^0$ TMDs in the H phase and smaller interlayer *o-e* interaction. This leads to the H phase tends to be electron depletion than the T phase. Figure 6(b) shows that, for the $d^0$ TiS$_2$, the intralayer *p-d* hybridization occurs between the occupied *p* orbital and empty *d* orbital, generating occupied and empty states that allow interlayer *o-o* and *o-e* interactions. For the $d^1$ NbS$_2$, the upper level from *p-d* hybridization is half-filled (which is empty for TiS$_2$). The half-filled level promotes interlayer *h-h* interaction. For $d^2$ MoS$_2$, the upper level is fully occupied, which facilitates interlayer multi-level interaction of more than one occupied level from each layer, namely, the *o-o(m)* interaction.

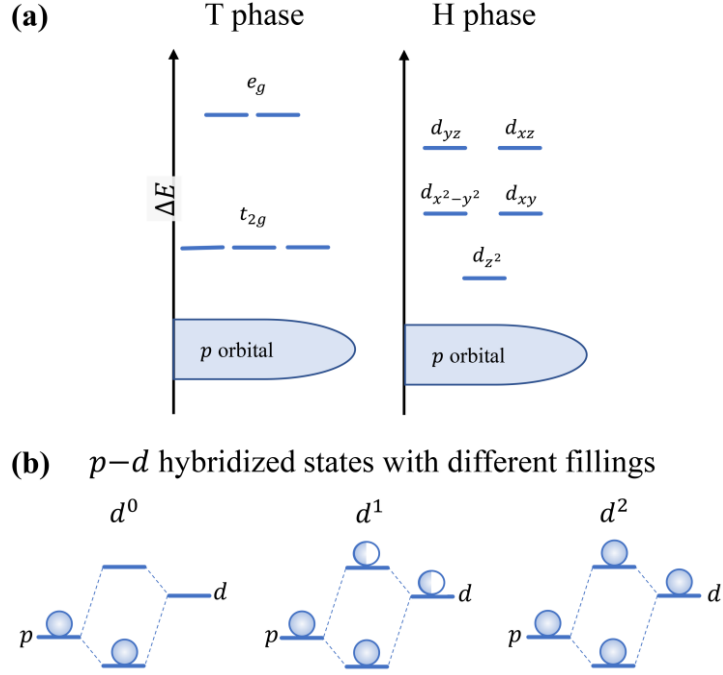

**FIG. 6. Schematic diagram of energy level arrangement in T and H phases (a), and the *p-d* hybridized states with different number of *d*-electrons filling (b).**

## IV. CONCLUSIONS

In summary, we have provided a unified understanding of the different ICDR behaviors in different TMDs ($d^0$ TiS$_2$, $d^1$ NbS$_2$, and $d^2$ MoS$_2$) with T or H phases based on the four types of interlayer QCB interactions. For $d^0$ TiS$_2$, the difference in ICDR between T and H phases arises from the competition of interlayer *o-o* and *o-e* interactions. Through DFT calculations and the two-center two-level model, we found that the *o-o* interaction depletes electrons ($P_{12}^{o,o}$) at the middle region of the vdW gap, due to its net antibonding character, while the *o-e* interaction accumulates electrons ($P_{12}^{o,e}$) via bonding interaction. For TiS$_2$, the $p_z$ orbital appears in both the VB and CB, which leads to a competition between *o-o* and *o-e* interactions. The difference between T and H phases is that: for the T phase, the $p_z$ orbital weight of CB+2@Γ is almost equivalent to that of the $p_z$ orbital of VB-2@Γ, whereas for the H phase, the $p_z$ orbital weight of CB@Γ is smaller than that of the $p_z$ orbital of

VB-2@Γ, indicating that the interlayer *o-e* interaction is stronger in T phase than that in H phase. This explains why the T phase TiS$_2$ tends to accumulate electrons compared to the H phase. For $d^1$ NbS$_2$, the interlayer interaction between two half-filled levels (*h-h* interaction), promoting electron accumulation ($P_{12}^{h,h}$) further. This causes the $d^1$ NbS$_2$ to be more inclined to accumulate electrons at the interlayer region than the $d^0$ TiS$_2$ for both T and H phases. For $d^2$ MoS$_2$, we extend the two-level interaction to multi-level interaction between occupied levels [*o-o(m)* interaction], which considers more than one occupied energy levels each layer. The competition between electron-depleting $P_{12}^{o,o}(m)$ and electron-accumulating $P_{34}^{o,o}(m)$, which is driven by the interlayer multi-level interaction, results in complex ICDR behavior in the T phase MoS$_2$. In addition, for TMDs of different groups and with different structural phases, the transition metal atom resides in distinct coordination environments and possesses different numbers of *d*-electrons filling, which are related to the various ICDR behaviors. Ultimately, this work directly addresses the fundamental question of why certain TMDs exhibit electron accumulation, depletion, or complex patterns in the van der Waals gap by linking the material's intrinsic parameters—*d*-electron filling ($d^0$–$d^2$) and structural phase (T/H)—to the specific types and interplay of interlayer orbital interactions that become active. This unified understanding has laid a solid foundation for the ICDRs between vdW layered material layers and hence is helpful for the in-depth study of interlayer engineering of vdW materials.

## ACKNOWLEDGMENTS

This work was supported by the National Natural Science Foundation of China (Grant No. 12274111), the Central Guidance on Local Science and Technology Development Fund Project of Hebei Province (No. 236Z0601G), the Natural Science Foundation of Hebei Province of China (No. A2023201029), the Excellent Youth Research Innovation Team of Hebei University (No. QNTD202412), the Advanced Talents Incubation Program of the Hebei University (Grant No. 521000981390), the Scientific Research and Innovation Team of Hebei University (No. IT2023B03), the Postgraduate's Innovation Fund Project of Hebei University (Grant No. HBU2026BS020), and the high-performance computing center of Hebei University.

## DATA AVAILABILITY

The data that support the findings of this article are openly available [45].

*Supplemental Material of:*

# On the origin of diverse interlayer charge redistribution in transition-metal dichalcogenides

Yu-Meng Gao, Nie-Wei Wang, Shi-Xuan Yuan, Wen-Xin Xia, Jiang-Long Wang*, and Xing-Qiang Shi*

Key Laboratory of Optic-Electronic Information and Materials of Hebei Province, Hebei Research Center of the Basic Discipline for Computational Physics, College of Physics Science and Technology, Hebei University, Baoding 071002, P. R. China

*E-mails: jlwang@hbu.edu.cn, shixq20hbu@hbu.edu.cn

**Contents** (Figures, Notes and Tables)**:**

Figure S1. ICDRs of T phase $TiS_2$ calculated with different density functionals;

Figure S2. The geometric structure of the T and H phases of transition metal dichalcogenides;

Figure S3. Band structure evolution from monolayer (1L) to bilayer (2L) of T and H phases $TiS_2$;

Note S1. The overlap populations ($P_{12}$);

Note S2. The meaning of $P_{12}^{o,o}(m)$ and $P_{34}^{o,o}(m)$ in the interlayer four-level interaction, which includes Fig. S4;

Table SI. Calculated energy difference (per formula unit) between T and H phases for bulk $TiS_2$, $NbS_2$, and $MoS_2$;

Table SII. Integrated electron accumulation/depletion at the middle of the van der Waals gap for $TiS_2$ and $NbS_2$ in T and H phases;

Table SIII. Interlayer S⋯S and Ti⋯S distances of $TiS_2$ in T and H phases.

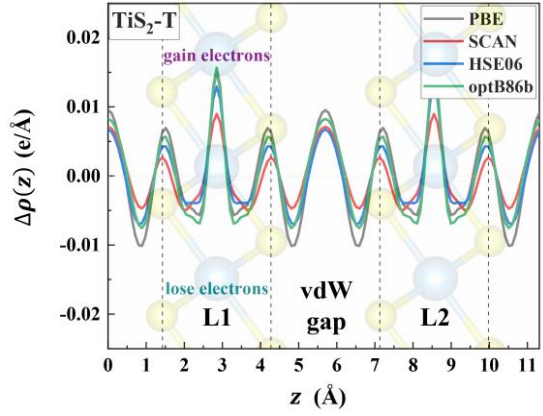

FIG. S1. ICDRs from the plane-averaged electron density redistribution along the $z$-direction, $\Delta\rho(z) = \rho(\text{bulk}) - \rho(\text{monolayers})$, for T phase $TiS_2$ calculated with different density functionals.

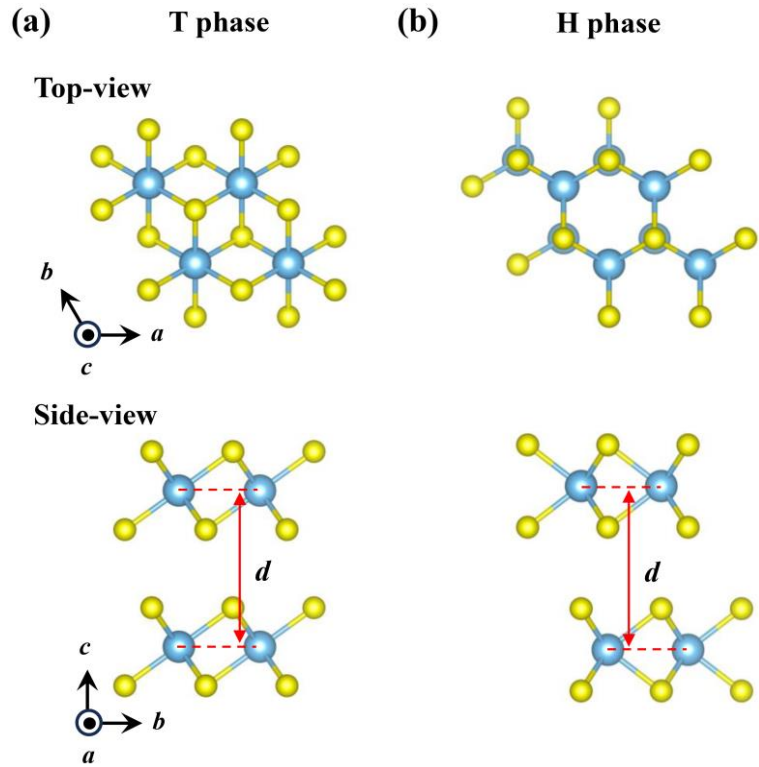

Fig. S2. Side- and top-views of transition metal dichalcogenides in T and H phases with an interlayer separation, $d$, as labeled in side-views. Yellow balls represent sulfur atoms, and blue balls represent transition metal atoms.

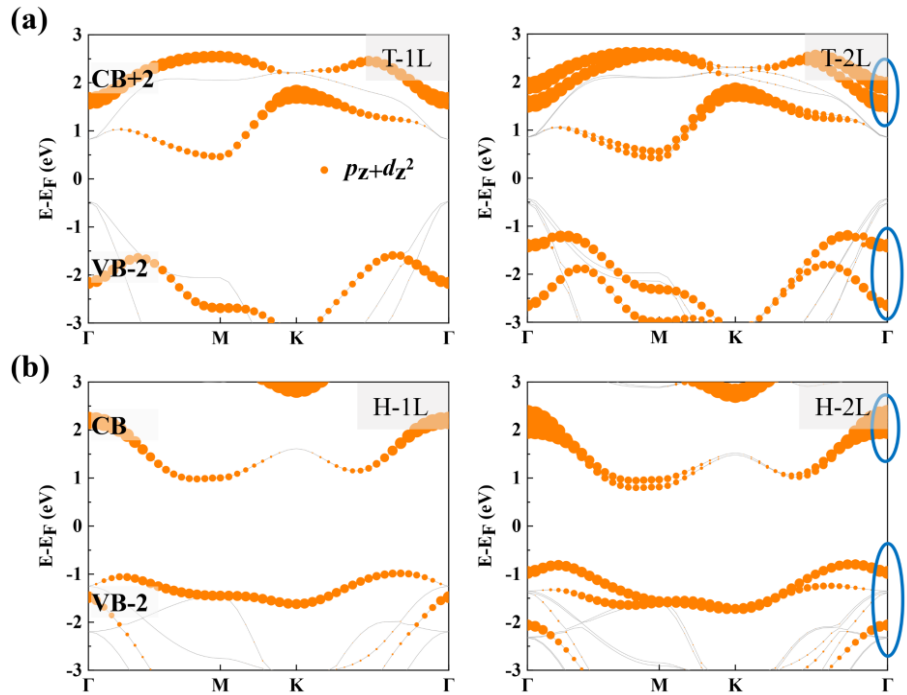

Fig. S3. Band structure evolution from monolayer (1L) to bilayer (2L) of T and H phases TiS$_2$. The band structures are projected to the out-of-plane $p_z$ and $d_{z^2}$ orbitals; the dot sizes are proportional to the magnitude of projection. In 1L (left panels), the bands labeled with CB+2/VB-2, CB/VB-2 indicate that the $p_z+d_{z^2}$ projections are larger at Γ point for these bands, and the corresponding energy splitting at Γ point caused by interlayer interaction in 2L (right panels) are indicated with ellipses.

## Note S1. The overlap populations ($P_{12}$)

In Figs. 1(a-c) of the main text, each molecular orbital ($\psi_i$, $i = 1, 2$), or crystal orbital for a given $k$ point, contains $n_i$ electrons ($n_i = 2, 0$ or $1$). Then, by the linear combination of atomic orbitals (LCAO) [1, 2],

$$\psi_i = \sum_\mu c_{\mu i} \chi_\mu \quad (\mu = 1, 2), \tag{S1}$$

$$P_{\mu\nu} = P_{12} = \sum_i 2n_i c_{1i} c_{2i} S_{12}, \quad (i = 1, 2) \tag{S2}$$

where $\mu, \nu$ are the atomic-orbital (or energy level) indices, $c_{1i}, c_{2i}$ are expansion coefficients of LCAO, and $P_{12}$ ($S_{12}$) is the overlap population (overlap integral) between atomic orbitals. $\psi$ and $\chi$ satisfies $\langle\psi_i|\psi_i\rangle = 1$, $\langle\chi_\mu|\chi_\mu\rangle = 1$ and $\langle\chi_\mu|\chi_\nu\rangle = S_{\mu\nu}$. Then, for the interaction between the same orbitals, or homo bilayers of van der Waals layered materials, for the bonding state $\psi_2$ in Fig. 1(a) of the main text, from Eq. (S1) one obtains

$$c_{12} = c_{22} = [2(1 + S_{12})]^{-\frac{1}{2}}, \tag{S3}$$

and for the antibonding state $\psi_1$,

$$c_{11} = -c_{21} = [2(1 - S_{12})]^{-\frac{1}{2}}. \tag{S4}$$

Substitute Eqs. (S3) and (S4) into Eq. (S2), one gets the overlap population.

For the overlap population in Fig. 1(a) of the main text between two fully-occupied orbitals (the *o-o* interaction), it is:

$$P_{12}^{o,o} = \frac{2S_{12}}{1 + S_{12}} - \frac{2S_{12}}{1 - S_{12}} = \frac{-4(S_{12})^2}{1 - (S_{12})^2} < 0. \quad \text{[net electron depletion]} \tag{S5}$$

The first term to the right of the first equal sign, $\frac{2S_{12}}{1+S_{12}}$, is the bonding state [$\psi_2$ in Fig. 1(a) of the main text] contribution to electron accumulation in the middle region; and the second term, $-\frac{2S_{12}}{1-S_{12}}$, is the antibonding state [$\psi_1$] contribution to electron depletion. The overall effect (or net effect) is $\frac{-4(S_{12})^2}{1-(S_{12})^2}$, namely, an overall negative overlap population occurs, which corresponds to a net electron depletion.

Also, for Fig. 1(c) of the main text, from Eq. (S2) with $n_i = 1$ for half-filled levels, the overlap population is $P_{12}^{h,h} = \frac{S_{12}}{1+S_{12}}$, namely, electron accumulation in the middle region.

In Fig. 1(b) of the main text, for the overlap population of occupied-empty interaction [1]:

$$P_{12}^{o,e} = \frac{4t_{12} S_{12}}{E^o - E^e} > 0, \quad \text{[electron accumulation]} \tag{S6}$$

where $t_{12}$ is the hopping integral.

**Note S2. The meaning of $P_{12}^{o,o}(m)$ and $P_{34}^{o,o}(m)$ in the interlayer four-level interaction**

Under the interlayer four-level interaction, the ICDR arises from a competition between electron-depleting $P_{12}^{o,o}(m)$ and electron-accumulating $P_{34}^{o,o}(m)$. Here, the notation "$m$" denotes interlayer multi-level interaction, which means that in addition to the interlayer two-level interaction ($\chi_1$-$\chi_2$ interaction or $\chi_3$-$\chi_4$ interaction, both are net antibonding), there are also interlayer $\chi_1$-$\chi_4$ and $\chi_3$-$\chi_2$ interactions. For $P_{12}^{o,o}(m)$, the additional interlayer interaction induces upward energy shifts in orbitals $\psi_1$ and $\psi_2$ relative to the interlayer two-level interaction. Compared to the $\chi_1$-$\chi_2$ two-level interaction, this four-level interaction enhances the net antibonding character, resulting in greater electron depletion, of which the energy level corresponding to $\psi_1$ serves as the dominant energy level (with red line) [Fig. S4(a)]. For $P_{34}^{o,o}(m)$, the additional interactions cause downward energy shifts in orbitals $\psi_3$ and $\psi_4$, which may change the net antibonding character (from two-level interaction) to weakly bonding character (from four-level interaction), and leads to the weak electron accumulation, of which the energy level corresponding to $\psi_4$ serves as the dominant energy level (with red line) [Fig. S4(b)], as conformed by the DFT results in Fig. 5(e) of the main text.

For the overall four-level interaction of Fig. 1(d) in the main text, a competition between the two effects, $P_{12}^{o,o}(m)$ with strengthed antibonding and $P_{34}^{o,o}(m)$ with weak bonding, manifests an overall net antibonding character. And hence the top red line of the Fig. 1(d) in the main text represents the overall net antibonding of the four-level interaction.

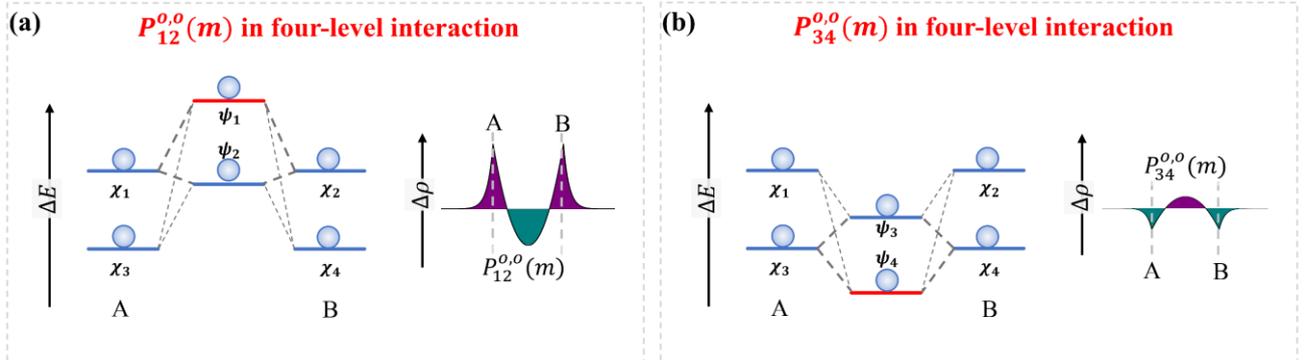

Fig. S4. Schematic diagrams of the interlayer four-level interactions for $P_{12}^{o,o}(m)$ and $P_{34}^{o,o}(m)$.

TABLE SI. Calculated energy difference (per formula unit) between T and H phases for bulk $TiS_2$, $NbS_2$, and $MoS_2$.

| E (eV/f.u.) | $TiS_2$ | $NbS_2$ | $MoS_2$ |
|---|---|---|---|
| $\Delta E$(T-H) | -0.460 | 0.143 | 1.023 |

TABLE SII. Integrated electron accumulation/depletion at the middle of the van der Waals gap for $TiS_2$ and $NbS_2$ in T and H phases. A positive/negative number means electron accumulation/depletion. $\Delta$(T−H) means the relative electron accumulation of T phase than H phase, and $\Delta$($NbS_2$−$TiS_2$) denotes the addition electron accumulation of $NbS_2$ than $TiS_2$ due to the additional interlayer interaction between half-filled levels (h-h interaction) in $NbS_2$.

| Electron transfer ($e$/cm$^2$) | $TiS_2$ | $NbS_2$ | $\Delta$($NbS_2$−$TiS_2$) |
|---|---|---|---|
| T phase | $2.19 \times 10^{11}$ | $2.36 \times 10^{12}$ | $2.14 \times 10^{12}$ |
| H phase | $-2.64 \times 10^{12}$ | $2.56 \times 10^{10}$ | $2.66 \times 10^{12}$ |
| $\Delta$(T−H) | $2.85 \times 10^{12}$ | $2.33 \times 10^{12}$ | N/A |

TABLE SIII. Interlayer S⋯S and Ti⋯S distances of $TiS_2$ in T and H phases, as well as the corresponding differences between interlayer S⋯S and Ti⋯S distances.

| | Phase | Interlayer S⋯S (Å) | Interlayer Ti⋯S (Å) | Difference (Å) |
|---|---|---|---|---|
| $TiS_2$ | T | 3.464 | 4.703 | -1.239 |
| | H | 3.642 | 4.597 | -0.955 |